\documentclass[twoside]{article}
\usepackage{amsmath}
\usepackage[psamsfonts]{amssymb}
\usepackage{cmmib57}
\usepackage[cmtip,arrow]{xy}
\usepackage{pb-diagram,pb-xy}
\usepackage{amsmath}
\usepackage[psamsfonts]{amssymb}
\usepackage{cmmib57}
\newcommand{\bPf}{\par\vspace*{-4pt}\indent{\sc Proof.}\enskip}
\newcommand{\ePf}{\medskip}
\def\QED{\hskip0.1em\hfill\null\ \null\nobreak\hfill\kern3pt\vbox{\hrule\hbox
   {\vrule\kern1pt\vbox{\kern1.7pt\hbox{$\scriptscriptstyle{QED}$}
    \kern0.2pt}\kern1pt\vrule}\hrule}}

\def\END{\hskip0.1em\hfill\null\ \null\nobreak\hfill\kern3pt\vbox{\hrule\hbox
   {\vrule\kern1pt\vbox{\kern1.7pt\hbox{$\,\,\,\vspace{5pt}$}
    \kern0.2pt}\kern1pt\vrule}\hrule}}
\newtheorem{theorem}{Theorem}
\newtheorem{lemma}{Lemma}
\newtheorem{corollary}{Corollary}
\newtheorem{proposition}{Proposition}
\newtheorem{remark}{Remark}
\newtheorem{definition}{Definition}
\newtheorem{example}{Example}
\newcommand{\bCd}{\bEq\begin{CD}}
\newcommand{\eCd}{\end{CD}\eEq}
\newcommand{\bcd}{\beq\begin{CD}}
\newcommand{\ecd}{\end{CD}\eeq}
\newcommand{\ben}{\begin{enumerate}}
\newcommand{\een}{\end{enumerate}}
\newcommand{\bEq}{\begin{eqnarray}}
\newcommand{\eEq}{\end{eqnarray}}
\newcommand{\beq}{\begin{eqnarray*}}
\newcommand{\eeq}{\end{eqnarray*}}
\newcommand{\bDf}{\begin{definition}\em}
\newcommand{\eDf}{\end{definition}}
\newcommand{\bLm}{\begin{lemma}}
\newcommand{\eLm}{\end{lemma}}
\newcommand{\bPr}{\begin{proposition}}
\newcommand{\ePr}{\end{proposition}}
\newcommand{\bTh}{\begin{theorem}}
\newcommand{\eTh}{\end{theorem}}
\newcommand{\bCr}{\begin{corollary}}
\newcommand{\eCr}{\end{corollary}}
\newcommand{\bRm}{\begin{remark}\em}
\newcommand{\eRm}{\end{remark}}
\newcommand{\bEx}{\begin{example}\em}
\newcommand{\eEx}{\end{example}}

\newcommand{\ie}{{\em i.e$.$} }

\newcommand{\R}{I\!\!R}

\DeclareMathOperator{\byd}{{\raisebox{.1ex}{:}{=}}}

\newcommand{\cE}{\mathcal{E}}

\newcommand{\cL}{\mathcal{L}}

\newcommand{\bK}{\boldsymbol{K}}

\newcommand{\bX}{\boldsymbol{X}}
\newcommand{\bY}{\boldsymbol{Y}}

\newcommand{\del}{\delta}
\newcommand{\eps}{\epsilon}

\newcommand{\lam}{\lambda}

\newcommand{\For}{{\Lambda}}

\newcommand{\Var}{{\mathcal{V}}}
\newcommand{\Thd}{{\Theta}}

\title{{\bf Symmetries of Helmholtz forms and globally variational dynamical forms
}}
\author{{\normalsize 
M. Palese and E. Winterroth}
\\{\footnotesize Department of Mathematics,
University of Torino}
\\{\footnotesize Via C. Alberto 10, 10123 Torino, Italy}\\ 
{\footnotesize e--mails: 
{\sc 
[marcella.palese, ekkehart.winterroth]@unito.it}}}
\date{}
\begin{document}

\maketitle

\begin{abstract}
Invariance properties of classes in the variational sequence
suggested to Krupka {\em et al.} the idea that there should exist a close correspondence between the 
notions of variationality of a differential form and invariance of its exterior derivative. 
It was shown by them that the  invariance of a closed Helmholtz form of a dynamical form is equivalent with local variationality of the Lie derivative of the dynamical form, so that the latter is locally  the Euler--Lagrange form 
of a Lagrangian. We show that the corresponding local system of Euler--Lagrange forms  is variationally equivalent to a global Euler--Lagrange form.

\medskip

\noindent {\bf 2000 MSC}: 55N30, 55R10, 58A12, 58A20.

\noindent {\em Key words}: variational sequence,
cohomology, symmetry, dynamical form, Helmholtz form.\end{abstract}

\section{Introduction}

Our general framework is the variational sequence \cite{Kru90} defined on a fibered manifold $\pi : \bY \to \bX$, with $\dim
\bX = n$ and $\dim \bY = n+m$; this approach can be encompassed within geometrical formulations of the calculus of variations
on fibered manifolds whereby the Euler--Lagrange operator appears as a morphism of an exact
sequence \cite{AnDu80,Kru90,Tak79,Tul77,Vin77}. The module in degree $(n+1)$ contains dynamical forms; a given equation is globally an Euler--Lagrange equation if its dynamical form is the differential of a Lagrangian and this is equivalent to the dynamical form being closed in the complex which is equivalent to Helmholtz conditions. The cohomolgy class of the dynamical form is then trivial.
Dynamical forms which are only {\em locally variational}, \ie which are closed in the complex and define a non trivial cohomology class, admit a system of local Lagrangians, one for each open set in a suitable covering, which satisfy certain relations among them. Analogously, $(n+2)$ dynamical forms (also called  here Helmholtz forms by an abuse of language) locally variational admit a system of local dynamical $(n+1)$ forms.
In the following we shall consider global projectable vector fields on a jet fiber manifold which are symmetries of locally variational Helmholtz forms. 

Let $h$ denote the horizontalization induced by the natural contact structure on jet prolongations of $\bY$ and denote by $d\ker h$ the sheaf generated by
the corresponding presheaf and set then $\Thd^{*}_{r}$ $\equiv$ $\ker h$ $+$
$d\ker h$.
The quotient sequence of the de Rham sequence by the naturally induced contact subsequence
\beq
0\arrow{e} \R_{\bY} \arrow{e} \dots\,\
\arrow[4]{e,t}{\cE_{n-1}}\,\ \ \For^{n}_r/\Thd^{n}_r
\arrow[3]{e,t}{\cE_{n}}\,\ \For^{n+1}_r/\Thd^{n+1}_r
\arrow[4]{e,t}{\cE_{n+1}}\,\ \ \For^{n+2}_r/\Thd^{n+2}_r
\arrow[4]{e,t}{\cE_{n+2}}\,\ \ \dots\,\ \arrow{e,t}{d} 0
\eeq
defines the {\em $r$--th order variational sequence\/}
associated with the fibered manifold $\bY\to\bX$. It turns out
that it is an exact resolution of the constant sheaf $\R_{\bY}$
over $\bY$.

The quotient sheaves (the sections of which are classes of forms modulo contact forms) in the variational sequence can be represented as sheaves $\Var^{k}_{r}$ of $k$-forms on jet spaces of higher order. In particular, currents are classes  $\nu\in(\Var^{n-1}_{r})_{\bY}$;
Lagrangians are classes $\lam\in(\Var^{n}_{r})_{\bY}$, while $\cE_n(\lam)$ is called a Euler--Lagrange form (being 
$\cE_{n}$ the Euler--Lagrange morphism); dynamical forms are classes $\eta\in(\Var^{n+1}_{r})_{\bY}$ and $\cE_{n+1}(\eta) \byd \tilde{H}_{d\eta}$ is a Helmohltz form (being $\cE_{n+1}$ the corresponding Helmholtz morphism).

Since the variational sequence is a soft
resolution of the constant sheaf $\R_{\bY}$ over $\bY$,  the cohomology of the complex of global sections is naturally isomorphic to both the \v Cech cohomology of  $\bY$ with coefficients in the constant sheaf $\R$ 
and  the de Rham cohomology $ H^k_{\text{dR}}\bY$ \cite{Kru90}.
Let $\bK^{p}_{r}\byd \text{Ker}\,\, \cE_{p}$.  We have the short exact sequence of sheaves
\beq 
0 \arrow{e}\bK^{p}_{r}\arrow[2]{e,t}{i}\, \, \Var^{p}_{r}\arrow[2]{e,t}{\cE_{p}}\, \
\cE_{p}(\Var^{p}_{r})\arrow{e} 0 \,.
\eeq 
In particular
$\cE_{n}(\Var^{n}_{r})$ is the sheave of Euler--Lagrange morphisms: for a global section $\eta\in(\Var ^{n+1}_{r})_{\bY}$ we have 
$\eta\in(\cE_{n}(\Var^{n}_{r}))_{\bY}$ if and only if $\cE_{n+1}(\eta)=0$, which are the Helmholtz conditions of local variationality.
Typically, a global inverse problem is to find necessary and sufficient conditions for such a locally variational $\eta$ to be globally variational.
The above exact sequence gives rise to the long exact
sequence in \v Cech cohomology 
\beq 
0 \arrow{e} (\bK^{p}_{r})_{\bY} \arrow{e}
(\Var^{p}_{r})_{\bY} \arrow{e}
(\cE_{p}(\Var^{p}_{r}))_{\bY} \arrow{e,t}{\del_{p}} H^{1}(\bY,
\bK^{p}_{r})\arrow{e} 0 \,. 
\eeq

\section{Locally  and globally variational dynamical forms}

As well known Noether Theorems relate  symmetries of a variational problem to conserved quantities; in order to make those theorems effective in the case of local systems, in  \cite{FePaWi10} we tackled the question what the most natural choice for {\em symmetries of the local variational problem } might be.
We use the concept of a {\em variational Lie derivative} operator $\cL_{j_{r}\Xi}$, defined for  any projectable vector field $(\Xi,\xi)$, which was inspired by the fact that the standard Lie derivative of forms with respect to a projectable vector field preserves the contact structure induced by the affine bundles $\pi^r_{r-1}$ (with $r\geq 1$) \cite{Kru73}. The variational Lie derivative is a local differential operator  by which symmetries  of Lagrangian and dynamical forms of any degree in the variational sequence, as well as corresponding Noether theorems,  can be characterized \cite{FPV02}. 
We notice that the variational Lie derivative sends a diagram of cochain complexes into a diagram of cochain complexes and thus defines an operator which acts on cohomology classes.

In order to study the obstruction to the existence of a variationally global equivalent to a local variational problem it is of fundamental importance to study how the  variational Lie derivative affects cohomology classes.
It is notewhorty that, independently from the fact that $\Xi$ be a dynamical form symmetry or not, {\em the variational Lie derivative trivializes cohomology classes} \cite{PaWi11}.
In fact, by linearity and resorting to the naturality of the variational Lie derivative we have
\beq
\eta_{\cL_{\Xi}\lam_{i}} = \cE_{n} (\Xi_V \rfloor \eta_{\lam}) + \cE_{n} (d_H \eps_i )=  \cE_{n} (\Xi_V \rfloor \eta_{\lam}) = \cL_{\Xi}\eta_{\lam}\,.
\eeq
The result that $\cL_{\Xi}\eta_{\lam} =\cE_{n} (\Xi_V \rfloor \eta_{\lam})$ is very important for the cohomology since it implies that $\del (\cL_{\Xi}\eta_{\lam})= \del(\eta_{\cL_{\Xi}\lam_{i}})  = 0$ although $\del(\eta_\lam) \neq 0$. From this we see that the variational Lie derivative enables us to transform non trivial cohomology classes to trivial cohomology classes associated with the variational Lie derivative of local presentations. On the other hand, since $\eta_{\cL_{\Xi}\lam_{i}} =\cE_{n} (\Xi_V \rfloor \eta_{\lam})$  we see that Euler--Lagrange equations of the local problem defined by $\cL_{\Xi}\lam_{i}$ are equal to Euler--Lagrange equations of the global problem defined by $\Xi_V \rfloor \eta_{\lam}$. Thus we have that the local problem defined by the local presentation $\cL_{\Xi}\lam_{i}$ is variationally equivalent to a global one \cite{PaWi11,PaWiGa11}.

This result holds true for local potentials of locally variational dynamical forms at any degree $p$ in the variational sequence. 
As it is well known, one of the results of the variational sequence theory, related to the inverse problem of the calculus of variations, states that a dynamical form $\eta$ is locally variational if and only if its Helmholtz 
form vanishes. Invariance properties of classes in the variational sequence
suggested to Krupka {\em et al.} the idea that there should exist a close correspondence between the 
notions of variationality of a differential form and invariance of its exterior derivative. 

Let us take into account symmetries of the Helmholtz form. It is well known \cite{KKPS10} that the  invariance of a closed Helmholtz form $\zeta_{\eta_i}$, \ie $\cL_{\Xi}\zeta_{\eta_i}$ is equivalent with local variationality of the Lie derivative $\cL _{\Xi}\eta_i$, \ie $\zeta_{\cL _{\Xi}\eta_i}=0$ meaning that the dynamical form 
$\cL _{\Xi}\eta_i$ is locally  the Euler--Lagrange form 
of a Lagrangian.
In the following we prove that a stronger result holds true; more precisely we prove that the system of local Euler--Lagrange forms $\cL _{\Xi}\eta_i$ is variationally equivalent (in the sense that they have the same Helmholtz form) to a global Euler--Lagrange form.

\bPr 
Helmholtz conditions of the local problem $\cL_{\Xi}\eta_i$ are Helmholtz conditions for the global problem defined by $\Xi_V\rfloor\zeta_{\eta_i}$.
\ePr

\bPf
We apply the fact that the variational Lie derivative trivializes cohomology classes to the $(n+2)$ degree closed variational classes in the Krupka's sequence. Locally variational $(n+1)$ dynamical forms  are dragged by the variational Lie derivative to dynamical forms always globally variational. 
Analogously to what seen for locally variationally trivial Lagrangians, suppose $\cE_{n+2} (\zeta)=0$ (higher degree Helmholtz conditions) which implies that there exists a local system of Euler--Lagrange forms $\eta_i$. 
Let $\Xi$ be a symmetry of $\zeta_{\eta_i}$, \ie $\cL_{\Xi} \zeta_{\eta_i}=0$; then we have $\cE_{n+1}(\cL_{\Xi}(\eta_i )) =0$. 
In fact, since
\beq
\zeta_{\cL_{\Xi}\eta_{i}} = \cE_{n+1} (\Xi_V \rfloor \zeta_{\eta})= \cL_{\Xi}\zeta_{\eta}\,,
\eeq
\ePf
the result is a straightforward consequence of Noether's theorem and of the naturality of $\cL$ \cite{FPV02} .

\subsection*{Acknowledgements}

Research  partially supported by GNFM of INdAM; E.W. is also supported by MIUR research grant {\em  Sequenze variazionali e Teoremi di Noether etc.}.


\end{document}